\documentclass[conference]{IEEEtran}
\ifCLASSINFOpdf
   \usepackage[pdftex]{graphicx}
   \graphicspath{{figures/pdf/}}
   \DeclareGraphicsExtensions{.pdf,.jpeg,.png}
\else
\fi
\begin{document}
\title{Detecting Post-Stroke Aphasia Via Brain Responses to Speech in a Deep Learning Framework}
\author{\IEEEauthorblockN{Pieter De Clercq\IEEEauthorrefmark{1}\textsuperscript{1}, Corentin Puffay\IEEEauthorrefmark{1}\textsuperscript{1,}\textsuperscript{2}, Jill Kries\textsuperscript{1}, Hugo Van Hamme\textsuperscript{2},\\Maaike Vandermosten\textsuperscript{1}, Tom Francart\textsuperscript{1} and Jonas Vanthornhout\textsuperscript{1}}
\IEEEauthorblockA{\IEEEauthorrefmark{1}Shared first authors\\
\textsuperscript{1}KU Leuven, dept. Neurosciences, ExpORL, Leuven, Belgium\\
\textsuperscript{2}KU Leuven, dept. Electrical engineering (ESAT), PSI, Leuven, Belgium\\
Correspondence: pieter.declercq@kuleuven.be, corentin.puffay@kuleuven.be}}
\maketitle

\begin{abstract}
Aphasia, a language disorder primarily caused by a stroke, is traditionally diagnosed using behavioral language tests. However, these tests are time-consuming, require manual interpretation by trained clinicians, suffer from low ecological validity, and diagnosis can be biased by comorbid motor and cognitive problems present in aphasia. In this study, we introduce an automated screening tool for speech processing impairments in aphasia that relies on time-locked brain responses to speech, known as neural tracking, within a deep learning framework. We modeled electroencephalography (EEG) responses to acoustic, segmentation, and linguistic speech representations of a story using convolutional neural networks trained on a large sample of healthy participants, serving as a model for intact neural tracking of speech. Subsequently, we evaluated our models on an independent sample comprising 26 individuals with aphasia (IWA) and 22 healthy controls. Our results reveal decreased tracking of all speech representations in IWA. Utilizing a support vector machine classifier with neural tracking measures as input, we demonstrate high accuracy in aphasia detection at the individual level (85.42\%) in a time-efficient manner (requiring 9 minutes of EEG data). Given its high robustness, time efficiency, and generalizability to unseen data, our approach holds significant promise for clinical applications.
\end{abstract}
\IEEEpeerreviewmaketitle

\section{Introduction}
The prevalence of aphasia, a language disorder primarily resulting from a stroke, is substantial, impacting over one-third of global stroke patients \cite{Gialanella2010}. Accurate and timely diagnosis of aphasia typically performed using behavioral language tests, is crucial for initiating effective therapy. However, language tests used in the clinic do not measure natural, daily-life language skills, as they consist of artificial tasks that target isolated language components (i.e., phonetic, semantic, or syntactic processes) \cite{Pasley2013}. Furthermore, these tests are time-consuming in administration and post-hoc interpretation by a trained clinician and can be biased by comorbid motor and cognitive problems present in over 80\% of individuals with aphasia (IWA) \cite{ElHachioui2014, Rohde2018}.

To address these limitations, electroencephalography (EEG) studies have explored biomarkers of impaired language functioning in aphasia. EEG-based neural tracking of natural speech measures time-locked brain responses to acoustic (e.g., the speech envelope) and linguistic (providing the semantic value carried by a word or a phoneme) \cite{Gillis2022} speech representations. Several methods are used to relate the neural signals to speech, for example using backward models that reconstruct the speech features from the neural signals \cite{Crosse2016} or models that attempt to discriminate time-aligned from misaligned speech segments (i.e., the match-mismatch paradigm) using the neural signals \cite{DeCheveigne2021}. Recent findings using linear models showed decreased neural tracking of the speech envelope and segmentation features (i.e., phoneme and word onsets) in IWA compared to healthy controls, but no group differences were observed for linguistic speech tracking \cite{Kries2023}. Focusing on the temporal envelope of speech, a follow-up study using nonlinear mutual information analyses demonstrated that primarily low-frequency envelope tracking (delta and theta band) is impaired in IWA \cite{DeClercq2023aphasia}, which carries cues for detecting and segmenting lexical units of speech (i.e., syllables, words, phrases and sentences) and prosody \cite{Giraud2012}. Using neural envelope tracking measures as input, classification algorithms reliably detected aphasia at the individual level \cite{DeClercq2023aphasia}. Together, these results show promise for neural tracking as an automated assessment tool for natural speech abilities in aphasia.

 A critical drawback in previous aphasia studies is that the applied techniques (i.e., linear models/mutual information) require participant- and stimulus-specific training data. Here, we test the potential of more powerful deep neural networks to generalize to unseen data and capture speech processing impairments in IWA, aiming to build a more robust and generalizable screening tool for aphasia. Specifically, we pre-train convolutional neural networks (CNNs) on acoustic (envelope), segmentation (phoneme/word onsets), and linguistic speech representations in a large independent sample of healthy individuals with intact speech processing. Subsequently, we evaluate the pre-trained CNNs on individual data from IWA and healthy age-matched controls. All models' performance for each speech feature is computed and provided as input to a nonlinear support vector machine (SVM) classifier trained to detect aphasia at the individual level. Furthermore, we investigate the contribution of each feature and the minimally required recording length for accurate aphasia detection. 

\section{Methods}
\subsection{Participants}
The presented sample is identical to previous work in our lab \cite{DeClercq2023aphasia}, comprising 26 IWA (seven female participants, mean age=72, std=15 y/o) in the chronic phase ($>$ 6 months after stroke) and 22 neurologically healthy controls (seven female participants, mean age=72, std=7 y/o), all Flemish-native speakers. IWA suffered from a left-hemispheric or bilateral stroke and were diagnosed with aphasia in the acute stage post-stroke. IWA and healthy controls both performed standardized clinical tests for aphasia at the time of study participation, with IWA displaying worse performance \cite{DeClercq2023aphasia}. Although 27\% of IWA did not score below threshold to be categorized as aphasia on either of these tests, these patients had an extended language deficit documentation and followed speech-language therapy at the time of study participation. Hence, they were still labelled as aphasia. For more details on recruitment strategy and the aphasia sample, we refer to \cite{DeClercq2023aphasia}. The study was approved by the ethical committee UZ/KU Leuven (S60007), and all participants gave written consent before participation. 

During the EEG experiment, participants were instructed to listen to a 25-minute-long story, "De Wilde Zwanen", written by Christian Andersen and narrated by a female Flemish-native speaker. For each subject, we used 20 minutes of the entire recording, as we experienced data loss for 1 participant in the last five minutes of recording. EEG data was recorded using a 64-channel Active-Two EEG system (BioSemi, Amsterdam, Netherlands) at 8192~Hz sampling rate.

\subsection{Signal processing and feature extraction}
\subsubsection{EEG data processing}
We performed pre-processing steps on the EEG data, including downsampling to 512~Hz, independent component analysis to remove artifacts, common average-referencing and filtering in frequency bands of interest: delta (0.5-4~Hz), theta (4-8~Hz), alpha (8-12~Hz), beta (12-25~Hz) and a broad (0.5-32~Hz) band. We used a Least Squares filter of order 2000 with 10\% transition bands (transition of frequencies 10\% above the lowpass filter and 10\% below the highpass) and compensation for the group delay. Furthermore, the EEG signals were z-score normalized and downsampled to 64~Hz.

\subsubsection{Speech features}
We extracted the speech features from the stimulus to investigate neural tracking at the acoustic (\textit{5 features}), segmentation (\textit{2 features}), and linguistic (\textit{4 features}) level. All features were downsampled to 64~Hz to be synchronized with the EEG signal. For detailed information about the implementation of features, we refer to our recent publications using the same feature sets \cite{Puffay_2023_ling, Kries2023, DeClercq2023aphasia}.

\begin{itemize}
  \item \textit{Acoustic features: }We extracted the temporal \textit{envelope} of speech in the \textit{delta}, \textit{theta}, \textit{alpha}, \textit{beta} and \textit{broad} band (reportedly involved in different stages of speech processing \cite{Giraud2012}), following the procedure used in our prior aphasia work \cite{DeClercq2023aphasia}. The envelopes were filtered to the frequencies of interest using the same Least Squares filter as applied on the EEG.
  \item \textit{Segmentation features: }The \textit{onsets} of \textit{phonemes} and \textit{words}, were extracted from the stimulus. The resulting features were one-dimensional vectors identifying the presence of a word (or a phoneme) onset with an impulse of unit magnitude, and zero otherwise.
  \item \textit{Linguistic features: } The linguistic features are encoded identically to the segmentation features, except for the unit magnitude that is replaced with a value reflecting the information carried by the word or the phoneme.
  At the phoneme level, we computed: (1) \textit{Phoneme surprisal:} reflects the probability of each phoneme given the preceding phonemes in the same word, and (2) \textit{Cohort entropy:} reflects the degree of competition among possible words created from the phoneme of interest. At the word level, we computed (3) \textit{Word frequency}: reflects how frequently a word occurs in the language and, (4) \textit{Word surprisal}: reflects the probability of each word given the preceding words in the same sentence.
\end{itemize}

\subsection{The match-mismatch task}
\begin{figure}[!t]
\centering
\includegraphics[width=3.1in]{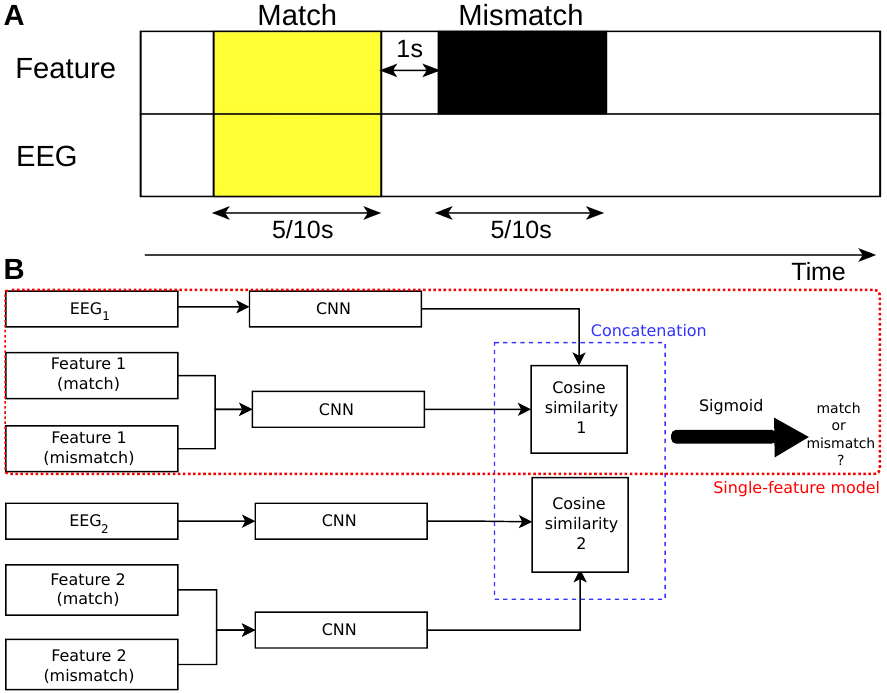}
\caption{\textbf{A. Match-mismatch task.} Finding which of the matched (yellow) or mismatched (black) segments is synchronized with the EEG segment. \textbf{B. Model architectures. }All models contain an EEG segment, a matched, and a mismatched segment of speech. They are fed to a CNN block, compared with cosine similarity, and passed through a last dense layer to select the matched segment. The single-feature architecture contains only the top stream, indicated with the red rectangle, while the dual-feature architecture includes the entire scheme.}
\label{fig:cnn}
\end{figure}

We trained all our models on the supervised match-mismatch task \cite{DeCheveigne2021}. As depicted in Figure \ref{fig:cnn}A, our models are trained to associate the EEG segment with the matched speech segment among two presented labeled speech segments. The matched speech segment is synchronized with the EEG, while the mismatched speech segment occurs 1~s after the end of the matched segment, following suggestions in prior work \cite{Puffay_2023_review}. We fixed the segment length to 5~s for the speech envelope and phoneme-based features. For word-based features, we chose 10~s segments to provide enough context to the models as suggested in \cite{Puffay_2023_ling}.

\subsection{Model architectures}

To evaluate the neural tracking of speech features, we used two CNN-based architectures depicted in Figure \ref{fig:cnn}B: (1) the single-feature model \cite{Accou2021ModelingTR, Puffay2022}, and (2) the dual-feature model \cite{Puffay2022, Puffay_2023_ling}. 
We used the single-feature architecture for the broadband envelope and the different frequency bands following up on \cite{Accou2021ModelingTR}. For the segmentation and linguistic features, we used the dual-feature model as proposed in \cite{Puffay_2023_ling}.
The single-feature model consists of two streams: one for the EEG segment, and one for the matched and mismatched speech segments. The EEG stream contains one spatial convolution followed by three temporal convolutions. The speech stream contains only the three temporal convolutions. The embedded representation of the EEG is compared using the cosine similarity with the embedded representation of the matched and mismatched speech segments, and the labels (i.e., match or mismatch) are predicted using a dense layer with a sigmoid activation function. The dual-feature model is the duplicated single-feature model's architecture, using two speech features instead of one, to select the matched segment. Both architectures are trained with the Adam optimizer and binary cross-entropy as the loss function. For more details, we refer to \cite{Accou2021ModelingTR, Puffay2022, Puffay_2023_ling}. We report in Table \ref{table_models} the features used for the given architectures.

\begin{table}[!t]
\renewcommand{\arraystretch}{1.3}
\caption{\textbf{Architectures used for each speech feature.}}
\label{table_models}
\centering
\begin{tabular}{|c||c|}
\hline
\textbf{Speech feature} & \textbf{Architecture}\\
\hline
Envelopes (N=5) & single-feature \\
\hline
2 x word onset & dual-feature \\
\hline
2 x phoneme onset & dual-feature \\
\hline
Word frequency + word onset & dual-feature \\
\hline
Word surprisal + word onset & dual-feature \\
\hline
Phoneme surprisal + phoneme onset & dual-feature \\
\hline
Cohort entropy + phoneme onset & dual-feature\\
\hline
\end{tabular}
\end{table}

\subsection{Model pre-training and evaluation on the match-mismatch task}

Each model we used was pre-trained on the match-mismatch task with a dataset consisting of 60 young native Flemish subjects listening to various speech materials involving different speakers, as used previously \cite{Puffay_2023_ling}. We trained, validated, and tested each model using an 80/10/10 split for each subject, following the method of our previous publications \cite{Accou2021ModelingTR, Accou_2021}. 
Subsequently, we evaluated our models on the aphasia and age-matched control participants. For each model, the evaluation metric is the average match-mismatch accuracy across all segments per participant. This resulted in 11 accuracy values per participant, one for each speech feature. We performed group comparisons on accuracy values using non-parametric Wilcoxon rank-sum tests, and p-values were adjusted for multiple comparisons employing the false discovery rate (FDR) correction with the Benjamini-Hochberg procedure \cite{Benjamini1995}.

\subsection{Aphasia classification}
The match-mismatch accuracies of all 11 models for each participant were used to classify the aphasia status of participants. In addition, age (which is matched between groups) was included as a feature, as it can affect neural tracking outcomes \cite{GillisKries}. We used a nonlinear radial basis function kernel SVM and performed a nested cross-validation approach. In the inner cross-validation, the C-hyperparameter and pruning were optimized (accuracy-based) and tested in a validation set using 5-fold cross-validation. Predictions were made on the test set in the outer loop using leave-one-subject-out cross-validation. We computed the receiver operating characteristic (ROC) curve and calculated the area under the curve (AUC), and further reported the accuracy, F1-score, sensitivity and specificity of the classifier.

Model interpretation was performed using SHAP (SHapley Additive exPlanations) analysis \cite{SHAP}. SHAP values quantify the impact of individual features on the SVM's classification, with positive values denoting features pushing predictions towards the positive class (aphasia), vice-versa for the negative class (healthy control), and values around 0 reflecting a limited feature impact.

Finally, we determined the required EEG recording length for accurate aphasia detection. We gradually increased the amount of EEG and speech data, ranging from 1~min to 20~min, with a step size of 1~min. For each duration and participant, we calculated the 11 match-mismatch accuracy values. We then repeated the SVM procedure for aphasia classification and computed the classification accuracy obtained for each recording length.

\section{Results}
\begin{figure}[!t]
\centering
\includegraphics[width=3.1in]{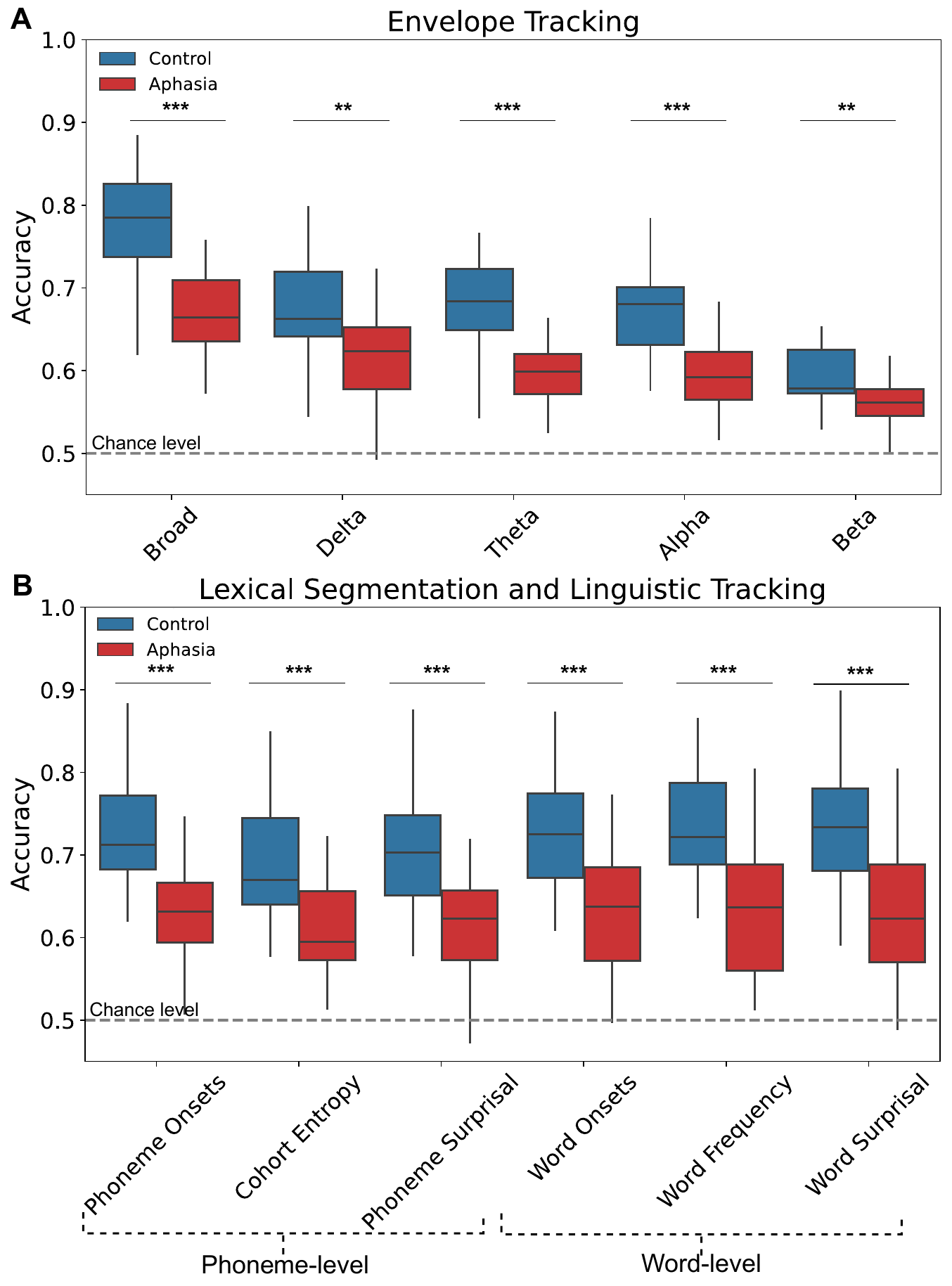}
\caption{\textbf{Match-mismatch accuracy values grouped for all features.} \textbf{A. } Accuracy values for all envelope-based features. \textbf{B. } Accuracy values for segmentation features (phoneme and word onsets) and linguistic features, grouped per level (phoneme/word level). Group comparisons were performed using Wilcoxon rank-sum tests. \textit{***: $p<.001$, **: $p<.01.$}}
\label{fig:acc}
\end{figure}

IWA exhibited statistically significant lower match-mismatch accuracies for all features, as depicted in Figure \ref{fig:acc}: broadband ($W=78$, $p<.001$), delta ($W=160$, $p=0.009$), theta ($W=75$, $p<.001$), alpha ($W=64$, $p<.001$), beta ($W=150$, $p=0.005$), phoneme onsets ($W=90$, $p<.001$), word onsets ($W=117$, $p<.001$), cohort entropy ($W=111.5$, $p<.001$), phoneme surprisal ($W=100$, $p<.001$), word frequency ($W=115.5$, $p<.001$), word surprisal ($W=117.5$, $p<.001$).

\begin{figure}[!t]
\centering
\includegraphics[width=3.1in]{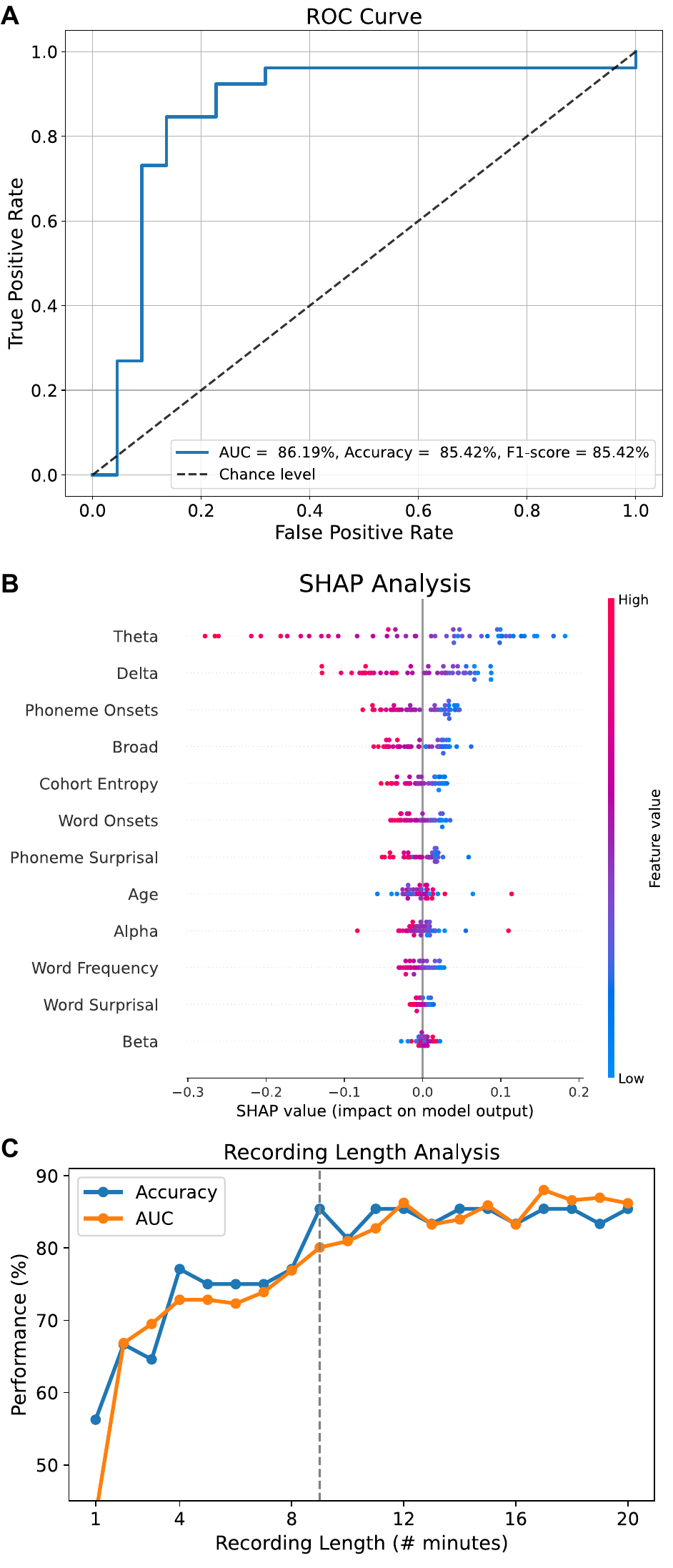}
\caption{\textbf{SVM performance and feature importance ranking.} \textbf{A. }ROC curve, plotting the false positive rate as a function of the true positive rate. \textbf{B. }SHAP analysis investigating the impact of each feature on the SVM's predictions. Features are ranked according to their impact. \textbf{C. }SVM performance as a function of EEG recording length. The 9-minute mark, achieving the same accuracy as the entire recording, is indicated in gray.}
\label{fig:svm}
\end{figure}

Next, we used individual participants' match-mismatch accuracy values for all 11 features, together with age, as input to an SVM to classify the aphasia status of participants using leave-one-out cross-validation. The SVM classified participants belonging to either group with an accuracy of 85.42\%, an F1-score of 85.42\% and an AUC of 86.19\%. The SVM had a sensitivity of 92.31\% and a specificity of 77.27\% for aphasia. Figure \ref{fig:svm}A displays the ROC curve. The SHAP analysis identified theta band envelope tracking as the strongest discriminative feature (see  Figure \ref{fig:svm}B).

Finally, Figure \ref{fig:svm}C displays the SVM classification accuracy as a function of the recording length. At the 9-minute mark, SVM accuracy approaches that of the full recording and consistently surpasses the 80\% accuracy and AUC benchmark thereafter.

\section{Discussion}
We reported decreased neural tracking in IWA compared to healthy controls for acoustic, segmentation, and linguistic speech features. The utility of neural tracking as an effective marker for aphasia was demonstrated using an SVM classifier which detected aphasia with high accuracy (85\%) and in a time-efficient manner (requiring 9 minutes of data). 

Decreased tracking of acoustic and segmentation features in IWA compared to healthy controls aligns with previous research on the same dataset \cite{Kries2023, DeClercq2023aphasia}. Leveraging the more complex and nonlinear architecture of CNN-based models, our approach successfully differentiated both groups in neural tracking of linguistic features as well, which linear models did not achieve \cite{Kries2023}. Using all speech features as input to an SVM, we obtained a small classification improvement of 2\% accuracy over a previous SVM classifier applied to the same dataset, which relied on mutual information-based tracking of the speech envelope only \cite{DeClercq2023aphasia}. It is plausible that SVM performance is at ceiling level in our sample, primarily consisting of mildly impaired IWA (27\% did not score below the clinical threshold for aphasia at time of study participation). Alternatively, the small improvement may mean that segmentation and linguistic features contribute only modestly over the envelope for aphasia classification. Nevertheless, this paper introduces a substantial enhancement compared to prior work in its clinical application potential, as CNN-based models demonstrate generalization to unseen data and speech materials. This eliminates the need for subject-specific training and enhances applicability to different settings.

Neural tracking of the low-frequency envelope (delta and theta) emerged as the most influential feature for aphasia classification (see Figure \ref{fig:svm}B). The envelope of speech is an essential cue for speech understanding \cite{Shannon1995}, encompassing cues for processing linguistic structures of speech \cite{Giraud2012}, and can be predicted from segmentation and linguistic speech features  \cite{Gillis2022}. This may explain the relatively low contribution of higher-level segmentation and linguistic features over the envelope for aphasia detection. Furthermore, the present approach including segmentation and linguistic features required slightly more data for optimal SVM performance compared to our previous results using only the envelope as input (9 vs. 7 minutes) \cite{DeClercq2023aphasia}. This suggests that CNNs may require more input data from linguistic features, aligning with prior findings using linear models \cite{Mesik2022}.

The presented approach exhibited remarkably high sensitivity to detect aphasia (92\%), surpassing the 73\% of the specific behavioral tests we administered in our study (see Methods and \cite{DeClercq2023aphasia}), suggesting a more sensitive screening tool for aphasia. However, before implementation in clinical settings, attention must be given to enhancing the specificity of our approach (currently at 77\%) which may require larger datasets. Furthermore, our approach must be validated against extensive behavioral test batteries, and its ability to capture impaired language components (i.e., specific problems with e.g. phonetic or semantic processes) must be assessed. Additionally, testing our approach on more severe IWA and in the acute post-stroke stage is essential, along with validation against stroke patients without aphasia (a critique that applies to behavioral tests as well, see \cite{Rohde2018}). Nevertheless, our work represents a significant step towards more automatic, objective, and time-efficient assessments of natural speech in aphasia.

\section{Conclusion}
The current study presents an automated screening tool for speech processing impairments in post-stroke aphasia using brain responses to natural speech in a deep learning framework. By leveraging deep neural networks trained on diverse speech materials and a large dataset of healthy individuals, which serves as a model for intact speech processing, our approach demonstrates the ability to generalize to unseen participants and to detect impaired speech processing in aphasia. With its high robustness, generalizability, and time-efficiency, our approach holds significant promise for clinical applications and opens new perspectives to studying various other language disorders with pre-trained CNNs.

\section*{Acknowledgment}
The authors would like to thank all participants. Thanks to Dr. Klara Schevenels for her assistance in the recruitment process and all individuals who helped with the data collection. Research was financially supported by the Research Foundation Flanders (PhD grant PDC: 1S40122N, PhD grant CP: 1S49823N, postdoc grant JV: 1290821, grant MV and TF: G0D8520N), the Luxembourg National Research Fund (FNR, AFR-PhD project JK: 13513810) and the KU Leuven Special Research Fund C24/18/099 (C2 project HVH and TF: 637424).

\bibliographystyle{IEEEtran}
\bibliography{IEEEabrv, bibliography}

\end{document}